\documentclass[a4paper,fleqn,usenatbib]{mnras}

\usepackage{newtxtext,newtxmath}

\UseRawInputEncoding

\usepackage[T1]{fontenc}
\usepackage{ae,aecompl}
\usepackage{pdflscape}
\usepackage[figuresright]{rotating}
\usepackage{tablefootnote}
\usepackage{longtable}
\usepackage{threeparttable}
\usepackage{caption}
\usepackage{comment}
\usepackage{adjustbox}

\usepackage{graphicx}	
\usepackage{amsmath}	
\usepackage{amssymb}	
\usepackage[LGRgreek]{mathastext}

\title[HCN emission in a candidate proto-brown dwarf]{HCN as a probe of the inner disk in a candidate proto-brown dwarf}

\author[Riaz, Thi, Machida]{
Riaz, B.,$^{1}$\thanks{E-mail: briaz@usm.lmu.de}
Thi, W.-F.,$^{2}$
Machida, M. N.$^{3}$
\\
$^{1}$  Universit\"{a}ts-Sternwarte M\"{u}nchen, Ludwig Maximilians Universit\"{a}t, Scheinerstra$\beta$e 1, 81679 M\"{u}nchen, Germany
\\
$^{2}$  Max-Planck-Institut f\"{u}r Extraterrestrische Physik, Giessenbachstrasse 1, D-85748 Garching, Germany 
\\
$^{3}$  Department of Earth and Planetary Sciences, Faculty of Sciences, Kyushu University, Fukuoka, Japan
}

\date{Accepted XXX. Received YYY; in original form ZZZ}

\pubyear{2024}

\begin{document}
\label{firstpage}
\pagerange{\pageref{firstpage}--\pageref{lastpage}}
\maketitle

\begin{abstract}


The detection of Keplerian rotation is rare among Class 0 protostellar systems. We have investigated the high-density tracer HCN as a probe of the inner disk in a Class 0 proto-brown dwarf candidate. Our ALMA high angular resolution observations show the peak in the HCN (3-2) line emission arises from a compact component near the proto-brown dwarf with a small bar-like structure and a deconvolved size of $\sim$50 au. Radiative transfer modelling indicates that this HCN feature is tracing the innermost, dense regions in the proto-brown dwarf where a small Keplerian disk is expected to be present. The limited velocity resolution of the observations, however, makes it difficult to confirm the rotational kinematics of this feature. A brightening in the HCN emission towards the core center suggests that HCN can survive in the gas phase in the inner, dense regions of the proto-brown dwarf. In contrast, modelling of the HCO$^{+}$ (3-2) line emission indicates that it originates from the outer pseudo-disk/envelope region and is centrally depleted. HCN line emission can reveal the small-scale structures and can be an efficient observational tool to study the inner disk properties in such faint compact objects where spatially resolving the disk is nearly impossible.




\end{abstract}


\section{Introduction} 
\label{intro}

Protoplanetary disks are the potential sites of planet formation, which makes them a key physical component to understand the transition from the late star-formation to the early planet-formation stages. The timescale of the formation of rotational disks is predicted to be during the very early star-formation stages when extremely young stars are in their embedded Class 0 phase (e.g., Andrews et al. 2018). Numerical simulations that consider non-ideal magneto-hydrodynamical effects, radiative transfer, turbulence, and initial conditions, such as, the misalignment between the magnetic field and the rotation axis, predict the formation of small disks of a few tens of au in Class 0 protostars (e.g., Machida et al. 2020). Molecular line interferometric observations have revealed rotational kinematics in a few Class 0 protostars, indicating the presence of a Keplerian disk, with radii of a few tens to a few hundred au (e.g., Aso et al. 2017; Hennebelle et al. 2017; Yen et al. 2017; Maret et al. 2020; Bergner et al. 2021; Chapillon et al. 2012). However, such observational evidence of rotational disks in Class 0 protostars has been very rare, mainly due to the fact that the disks are still deeply embedded in an infalling envelope, which makes it difficult to clearly distinguish the rotational kinematics from infalling motions (e.g., Guzmán et al. 2021; 2015; Qi et al. 2008).

Here, we present an investigation of a disk-like structure in a proto-brown dwarf candidate, Ser-emb 16, which is in its early Class 0 evolutionary stage (Riaz et al. 2018; 2024). Numerical simulations on brown dwarf formation predict sizes of $\leq$10 au for the inner Keplerian disk (e.g., Machida et al. 2009; van Zadelhoff et al. 2003), which is difficult to spatially resolve even in high angular resolution ALMA observations. An alternative way is to use high-density molecular line tracers that can probe the innermost densest regions in compact, dense proto-brown dwarfs. The HCN (3-2) transition line is one such tracer. At a critical density of the order of 10$^{7}$ cm$^{-3}$, it can probe the inner dense regions of a proto-brown dwarf where densities are predicted to be $>$10$^{6}$ cm$^{-3}$ (Machida et al. 2009).

Our target of interest is Ser-emb 16 that is located in the Serpens region ($d$=440$\pm$9 pc) (Ortiz-Le\'{o}n et al. 2017; Gong et al. 2021). The bolometric luminosity for Ser-emb 16 is 0.06$\pm$0.01 L$_{\odot}$ and the total (dust+gas) mass (central object mass and circumstellar material) is $\sim$28 M$_{Jup}$ (Riaz et al. 2024), which are well below the sub-stellar limits of $<$0.08 M$_{\odot}$ and $<$0.1 L$_{\odot}$ typically considered in brown dwarf evolutionary models. This YSO is therefore a strong proto-brown dwarf candidate. A detailed discussion on the present and final mass of Ser-emb 16 is presented in Riaz et al. (2024). Paper 1 (Riaz et al. 2024) presented the results in the ALMA continuum and HCO$^{+}$ (3-2) line. Here we present the results from ALMA HCN (3-2) line observations of Ser-emb 16. Previous single-dish IRAM 30m molecular line surveys have identified Ser-emb 16 to be chemically rich (Riaz et al. 2019a; 2022abc; 2023), with strong emission in the HCN (3-2) hyperfine lines (Riaz et al. 2018). We have followed-up with ALMA high angular resolution observations to investigate the origin of the HCN line emission in this proto-brown dwarf.

\section{Observations and Data Analysis}
\label{obs}

We observed Ser-emb 16 with ALMA in the HCN (3-2) line at 265.8865 GHz in Band 6. The observations were obtained in Cycle 8 (PID: 2021.1.00134.S) in September, 2022. The primary beam of the ALMA 12m dishes covers a field of $\sim$30 arcsec in diameter. Our chosen configuration was a maximum recoverable scale of $\sim$4 arcsec, and an angular resolution of $\sim$0.4 arcsec. The synthesized beam size of the observations is 0.47$^{\prime\prime} \times$0.38$^{\prime\prime}$. The bandwidth was set to 117.18 MHz, which provides a resolution of 244.14 kHz, or a velocity resolution of $\sim$0.3-0.4 km s$^{-1}$ at 265.8 GHz. One baseband was used for continuum observations where the spectral resolution was set to 976 kHz and the total bandwidth was 1875 MHz. We employed the calibrated data delivered by the EU-ARC. Data analysis was performed on the uniform-weighted synthesized images using the CASA software. We have used the CASA tasks {\it uvmodelfit} and {\it imfit} to measure the source size, peak position, position angle, and the line fluxes by fitting one or more elliptical Gaussian components on the continuum and HCN (3-2) line images.

 \begin{figure}
  \centering      
       \includegraphics[width=2.5in]{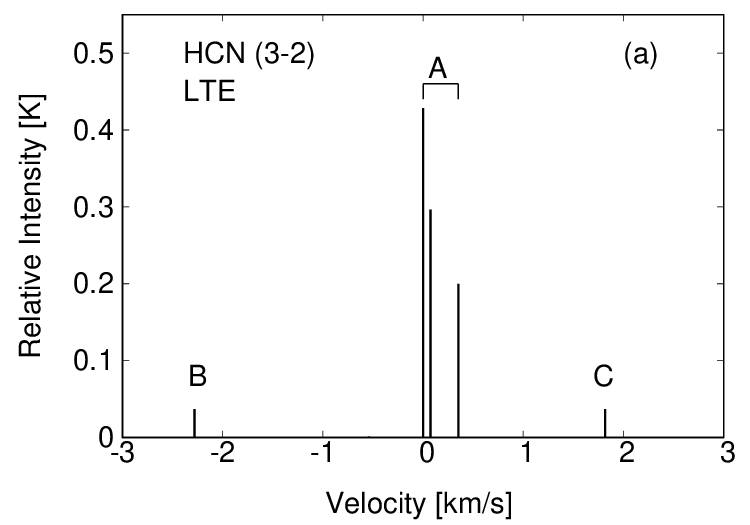}            
       \includegraphics[width=2.5in]{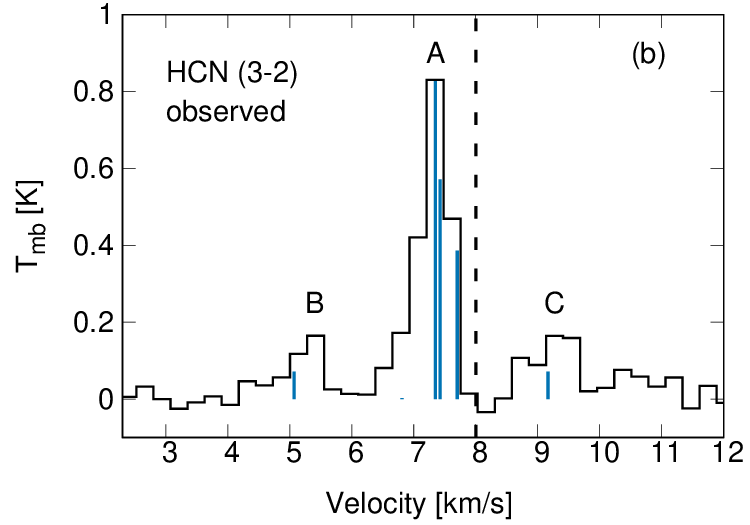}
           \caption{(a) The HCN hyperfine structure under LTE conditions. The main hyperfine components are labelled. (b) The ALMA HCN (3-2) spectrum for Ser-emb 16 (black line). The 1-$\sigma$ rms is $\sim$0.06 K. Dashed line is the source V$_{lsr}$. Blue lines mark the HCN hyperfine structure, as shown in the top panel, with the peak intensity matched for the A component. }
           \label{cont}
  \end{figure}

\section{Results}
\label{results}

\subsection{Anomalous hyperfine structure}

Figure~\ref{cont}b shows the observed HCN (3-2) spectrum collapsed over all velocity channels and measured over an area of 3x3 arcsecond that covers the full spatial scale of the observed structures (Fig.~\ref{mom0}). The HCN (3-2) spectrum shows hyperfine splitting. The hyperfine structure expected under optically thin LTE conditions is plotted in Fig.~\ref{cont}a. HCN (3-2) has six hyperfine components. The central bright feature (labelled `A') is a composite of four components that are not spectrally resolved, and the spectrum has the appearance of a strong central component with two satellite components (labelled `B' and `C'). Under optically thin LTE conditions, the relative weightings of the HCN (3-2) hyperfine lines are 0.037:0.926:0.037. The satellite hyperfine components `B' and `C' are thus expected to be of equal intensities, with the B/A and C/A line ratios equal to 0.04, assuming that the level populations of the hyperfine states are in LTE and hence proportional to the statistical weights. Ratios between 0.04 and 1 indicate anomalous sources and imply different hyperfine line ratios than expected due to various effects.

For the case of Ser-emb 16, the HCN-B and HCN-C components have an equal intensity of T$_{mb}$ = 0.16 K, and the HCN-A component is $\sim$5 times stronger with T$_{mb}$ = 0.83 K (Fig.~\ref{cont}b). This implies B/A and C/A line ratios of $\sim$0.2. We therefore find hyperfine anomalies that are related to non-equal excitation temperature among the hyperfine lines of a given multiplet. Radiative transfer effects of absorption or scattering of the photons emitted from the high-density core in the less dense outer envelope could produce such anomalies. Other factors such as line trapping effects or extra velocity components along the line of sight, could also produce higher or lower relative line intensity ratios of the hyperfine components than the expected values (e.g., Riaz et al. 2018). The anomalies are further discussed in Sect.~\ref{model}.


\subsection{Moment maps}
\label{mom-maps}

\begin{figure*}
  \centering      
       \includegraphics[width=1.85in,angle=90]{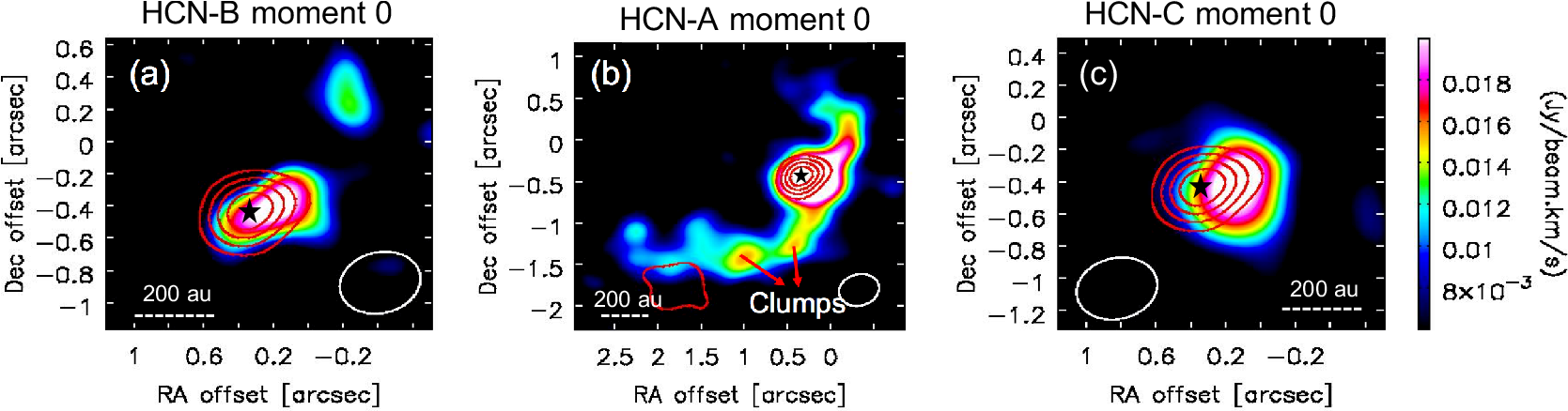}             
           \caption{The HCN moment 0 maps (colour maps) in the individual hyperfine components. Overplotted is the continuum emission (red contours). Star marks the Ser-emb 16 position.  }
           \label{mom0}
  \end{figure*}

Figures~\ref{mom0}abc show the moment 0 (integrated value of the spectrum) maps individually for the B ($\sim$4-6 km/s), A ($\sim$6.2-8.3 km/s), C ($\sim$8.3-10 km/s) hyperfine components in the HCN (3-2) line. The satellite HCN-B and HCN-C components (Fig.~\ref{mom0}ac) show a compact object that overlaps with the continuum emission for Ser-emb 16. In comparison, an extended ``spiral'' like structure originating from the proto-brown dwarf position is seen for the brightest HCN-A component, with some faint clumps within the spiral (Fig.~\ref{mom0}b). The brightest emission in all three components is thus seen originating from the proto-brown dwarf position.

\begin{figure*}
  \centering      
       \includegraphics[width=2in, angle=90]{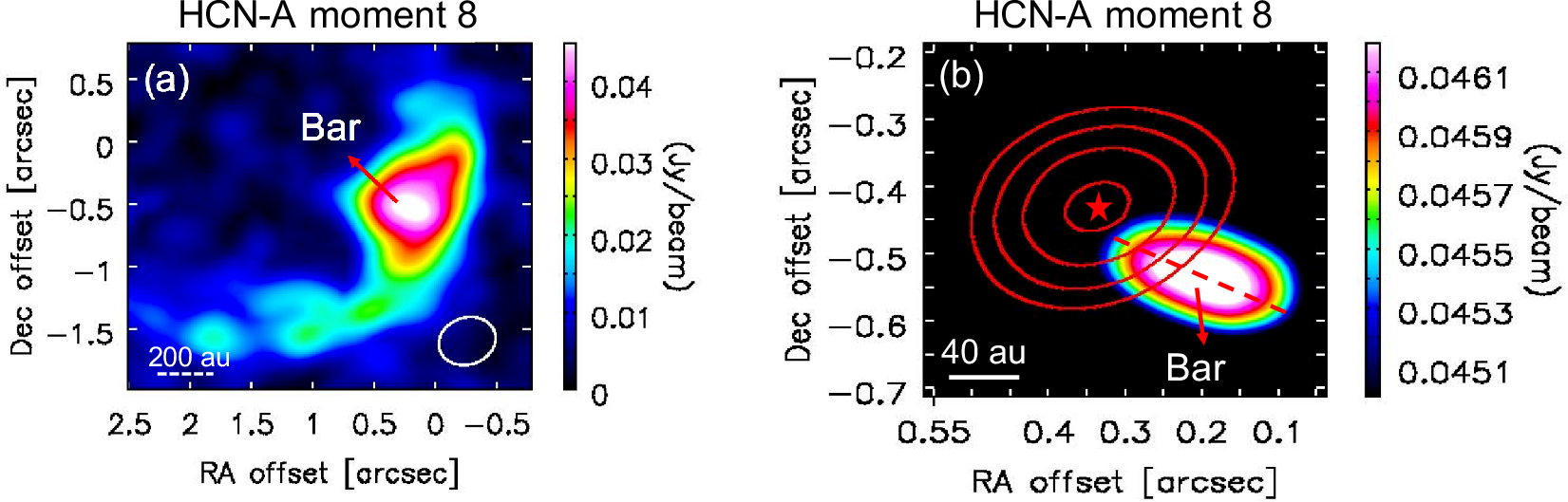}             
           \caption{The HCN moment 8 map in the brightest HCN-A component. Right panel (b) shows a zoomed-in view of the bar-like structure that shows the peak in the HCN emission. Also overplotted is the continuum emission (red contours). Star marks the Ser-emb 16 position. }
           \label{mom8}
  \end{figure*}

The moment 8 (maximum value of the spectrum) map in the HCN-A component (Fig.~\ref{mom8}ab) shows that the peak/maximum in the emission originates from a flattened ``bar'' like structure that lies very close ($\sim$0.1$\arcsec$ or $\sim$44 au) to the continuum peak position. The size of this peak emission region deconvolved by the beam is $\sim$0.11$\arcsec$ ($\sim$50 au) with a PA of 58.8$\degr$. This size was determined by performing a 2D Gaussian fit using the {\it imfit} task in CASA and restricting the fit to a region excluding the filamentary (or spiral) structure. The origin of this bar-like feature is explored in Sect.~\ref{model}.


The moment 0 map shows the integrated value of the spectrum, whereas the moment 8 map shows the maximum value of the spectrum. In Fig.~\ref{mom8}ab, for the case of the bar, we want to highlight the region of origin of the peak or maximum value of the spectrum, due to which moment 8 is preferred over moment 0 map. For the case of the spiral, moment 0 and 8 show the same results. Thus, moment 8 is necessary to highlight the bar but not to separate the bar and spiral. The distinction between the spiral and bar features is made visually. It is an intensity cut in the moment 8 map. To highlight the bar-like feature, we selected a range in intensities close to the peak value of $\sim$0.04 Jy/beam. For the spiral, the full range in intensities is shown. This indicates that we are seeing the highest intensities in the bar, likely because it is tracing the high-density regions in the proto-brown dwarf system.

The bar-like structure does not have similar strengths across all three components, otherwise this feature could be seen in the moment 8 map for the B and C components. These fainter components are barely detected at a $\sim$3-$\sigma$ level compared to A that is detected at a $>$10-$\sigma$ level. Since B and C are weakly detected, the bar is not visible in these components. The spiral structure is expected to have some contribution near the source position from the fainter B and C components. The extended part or the tail of the spiral as seen in the A component is at least a factor of $\sim$2 fainter than the emission seen close to the source position. It would thus be much more difficult to detect this extended tail of the spiral in the B and C components.

\section{Line radiative transfer modelling}
\label{model}

\begin{figure*}
  \centering      
       \includegraphics[width=2in, angle=90]{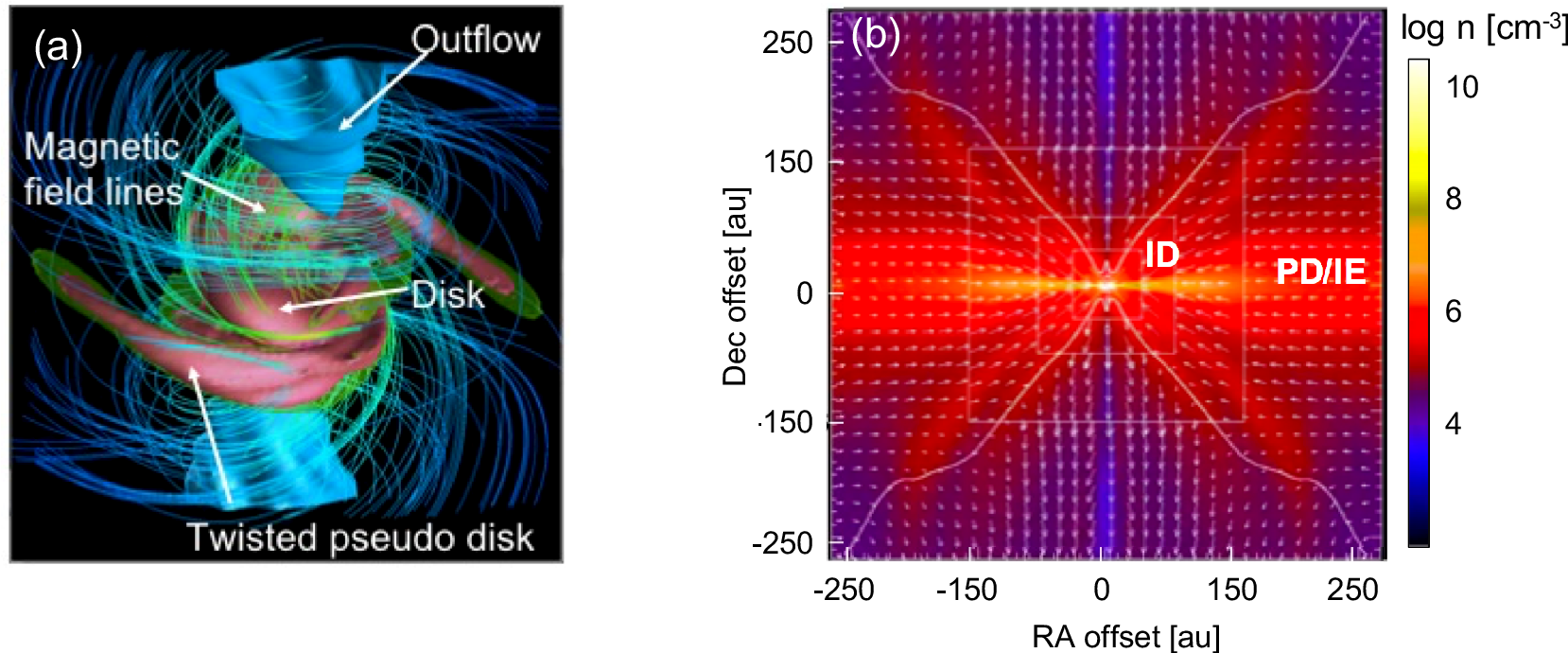}                 
           \caption{The physical structure of the Machida et al. (2020) model over a large ($\sim$1000 au) scale (left panel) and close to the central source (right panel). The labels `ID' and `PD/IE' indicate the inner disk and pseudo-disk/inner envelope, respectively. This model map is shown to be perfectly symmetric about the (0,0) position i.e. the location of the central object (Riaz et al. 2024).} 
           \label{sim}
  \end{figure*}

We consider the magnetohydrodynamical simulations presented in Machida, Hirano, \& Kitta (2020) of magnetized rotating clouds in which the rotation axis is misaligned with the global magnetic field by an angle ranging between 0$\degr$ and 90$\degr$. At large scales of $>$500 au, the pseudo-disk/envelope in this model is spatially stretched or extended due to the rotation of the proto-brown dwarf core in the presence of a strong magnetic field (Fig.~\ref{sim}a; Riaz et al. 2024). Close to the central source ($\leq$200 au), the simulations show distinct physical components of an inner Keplerian disk, a inner pseudo-disk/envelope, a high-velocity jet, and a low-velocity outflow, as shown in the generic model figure in  Fig.~\ref{sim}b. We have modified the simulations of the Machida et al. (2020) model. The modifications are described in Riaz et al. (2024).

\begin{figure*}
  \centering      
       \includegraphics[width=3in]{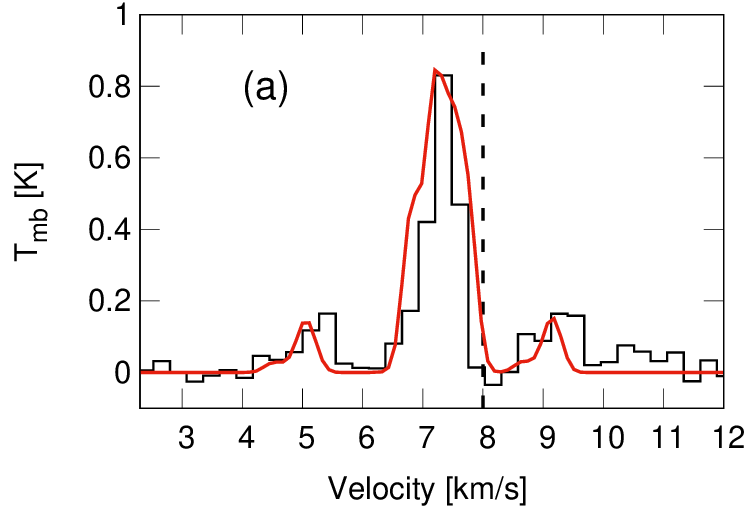}   \\
       \includegraphics[width=2.3in,angle=90]{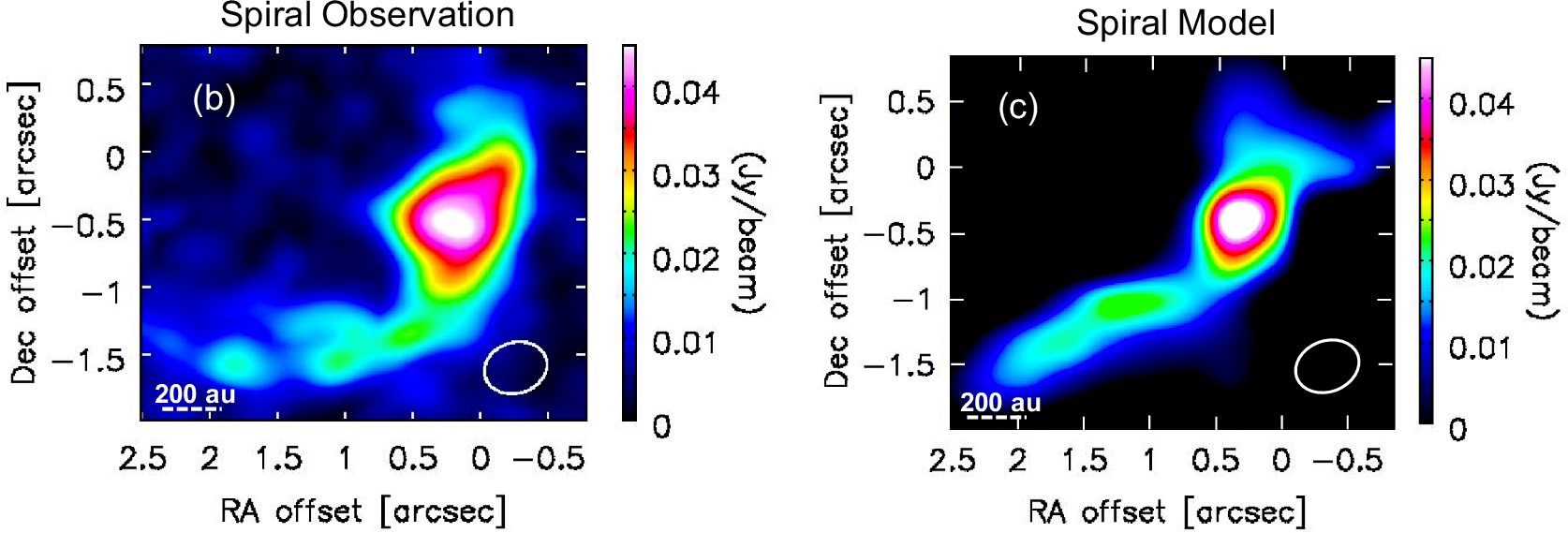}  \\    \vspace{0.2in}      
       \includegraphics[width=2.3in,angle=90]{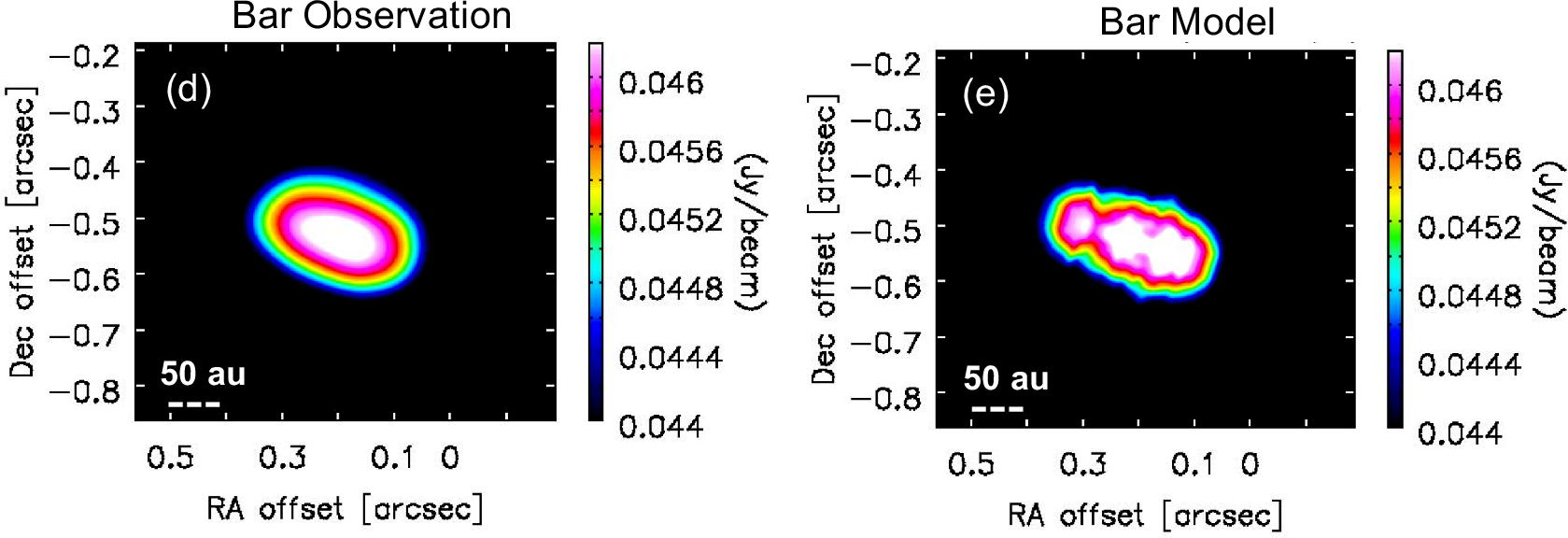}      
           \caption{Results from radiative transfer modelling of the HCN emission. (a) The best model fit (red line) to the observed (black line) HCN (3-2) spectrum. Panels (b) and (c) show the observed and model moment 8 map produced from the best-fit, respectively, over a large ($\sim$1000 au) scale.  Panels (d) and (e) show the observed and model moment 8 map produced from the best-fit, respectively, close to the proto-brown dwarf.  }
           \label{fit}
  \end{figure*}

We conducted line radiative transfer modelling following the method described in detail in Riaz et al. (2019b). The physical structure, i.e. the radial profiles of density, temperature, and velocity from the simulations is used as an input to the 3D non-LTE radiative transfer code MOLLIE (Keto et al. 2004). Synthetic HCN line profiles convolved by the ALMA beam size are compared with the observed spectrum. A reasonable fit to the strength and width of the observed line profile is reached from varying the abundance profile, the line width, the inclination angle of the system, and the angle between the rotation axis and the global magnetic field. For each synthetic spectrum, a reduced-$\chi^{2}$ value is computed to determine the goodness of fit. From the best line model fit (lowest reduced-$\chi^{2}$ value), model moment maps convolved by the beamsize are produced and compared with the observed maps. We find a model with the HCN molecular abundance relative to H$_{2}$ of (1-2)$\times$10$^{-8}$, an initial misalignment angle of 30$\degr$, and an inclination angle of 60$\degr$ provides the best-fit to the observed spectrum (Fig.~\ref{fit}a) and best matches the observed spiral and bar-like structures (Fig.~\ref{fit}b-e).



 \begin{figure}
  \centering          
       \includegraphics[width=2.7in]{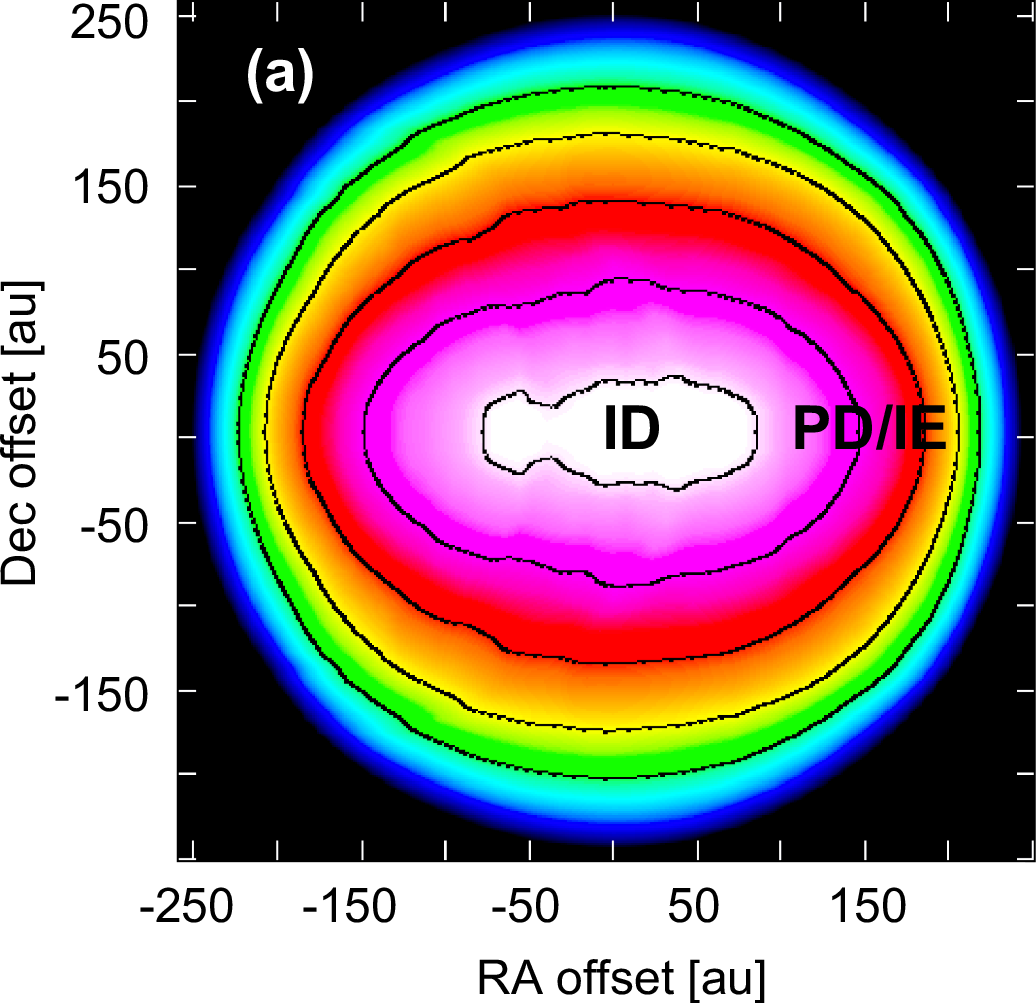}     \\  \vspace{0.2in} 
       \includegraphics[width=2.8in]{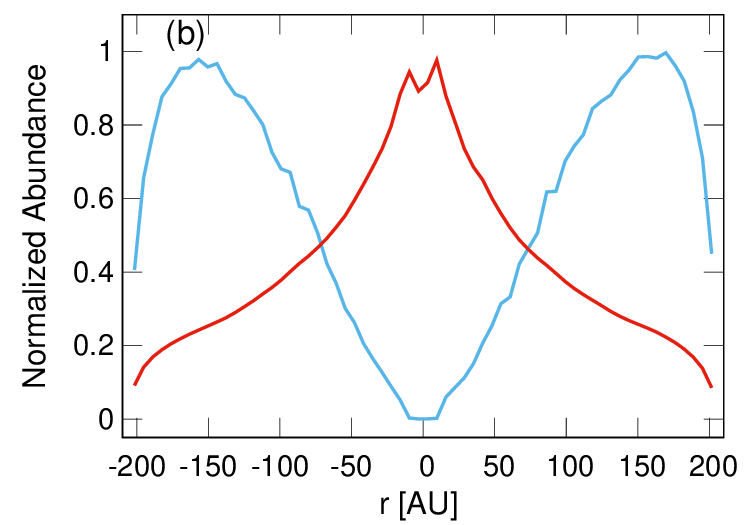}
           \caption{Results from radiative transfer modelling of the HCN emission. Panel (a) shows the model map at a edge-on inclination. The labels `ID' and `PD/IE' indicate the inner disk and pseudo-disk/inner envelope, respectively. This model map is shown to be perfectly symmetric about the (0,0) position i.e. the location of the central object. The contours are from 1-$\sigma$ to 5-$\sigma$ in steps of 1-$\sigma$. The 1-$\sigma$ rms is $\sim$40 mJy/beam. Panel (b) shows the normalized abundance (relative to H$_{2}$) profiles for HCO$^{+}$ (blue) and HCN (red) derived from the best-fit. The normalization constant is 1.7$\times$10$^{-8}$ and 5.0$\times$10$^{-8}$ for HCN and HCO$^{+}$, respectively. }
           \label{fit-2}
  \end{figure}

Modelling indicates that the spiral is tracing the pseudo-disk structure formed by the process of gravitational collapse with the rotation of an east-west angular momentum vector, which will cause the pseudo-disk to drag the magnetic field and finally connect it to the proto-brown dwarf (Fig.~\ref{fit}bc). Close to the central proto-brown dwarf, the model map shows a flattened, bar-like structure, similar in size to the observed bar (Fig.~\ref{fit}de). The origin of this structure can be seen more clearly in the edge-on (90$\degr$) model map (Fig.~\ref{fit-2}a). Here, HCN shows a brightening in emission in the inner most dense region within $\sim$100 au. The bar-like shape of this peak emission region suggests that HCN is tracing the high-density emission towards the inner disk (labelled `ID') in the proto-brown dwarf. Note that since the proto-brown dwarf system is viewed at a $\sim$60$\degr$ angle to the line-of-sight, we are seeing one-sided or asymmetric emission from the disk/pseudo-disk/envelope.

We note that MOLLIE will account for the splitting in the HCN hyperfine components and their relative optical depth effects (e.g., Loughnane et al. 2012; Mullins et al. 2014). The line radiative transfer model originally produces B/A and C/A hyperfine intensity ratios in HCN of $\sim$0.05, which is close to the LTE value of 0.04. Varying the radial profile of the abundance when modelling the observed spectrum produces a noticeable change in these hyperfine component ratios. For e.g., reducing the molecular abundance results in reducing the ratios. When modelling the spiral and the bar, the size of the region where the emission originates from was also constrained. When making the model moment maps from the best-fit to the spectrum, we had matched the intensity scale and the range in velocities in the model with the observations. Therefore, based on the modelling results, we can say that the hyperfine line ratios are truly anomalous. 

We have shown through modelling that the spiral and bar structures are tracing different physical components in the Ser-emb 16 system, namely, the spiral is likely tracing the extended/twisted pseudo-disk/envelope regions while the bar is likely tracing the inner Keplerian disk. Different physical regions imply different excitation conditions and different physical conditions, such as, the physical sizes, temperatures, densities, optical depths. Optical depth depends on the level population and the kinetic temperature effects. The effects of these are coupled and cannot be clearly distinguished. The anomalies seen in the HCN hyperfine structure can therefore be due to different hyperfine components tracing different physical regions. Such anomalies cannot be pinpointed to one particular effect.

Figure~\ref{fit-2}b shows the HCN abundance profile for the best model fit. There is an enhancement in the HCN abundance towards the core center, which suggests that HCN can survive in the gas phase in the inner, dense regions of the proto-brown dwarf. As shown in the Supplementary Material (Appendix C), the HCO$^{+}$ model moment 0 map generated from the best-fit indicates depletion in the central dense regions where a peak in the HCN emission is seen (Fig.~C1 in Appendix C). This is also seen in the abundance profile for HCO$^{+}$ best model fit (Fig.~\ref{fit-2}b) where the abundance is depleted towards the core center. The HCN and HCO$^{+}$ molecules thus provide a contrasting view of the internal structure in the proto-brown dwarf, such that, HCN can reveal the small-scale structures in regions where HCO$^{+}$ is depleted.

Our results are consistent with previous studies on molecular line observations of low-mass protostars, which have shown that the depletion of molecular species from the gas phase in the dense interiors of the cores is a highly selective process (e.g., Lee et al. 2004). Molecules like CO and HCO$^{+}$ disappear rapidly from the gas phase and suffer a significant drop in abundance towards the core center, while species like HCN and N$_{2}$H$^{+}$ survive much longer at high densities and are present in the gas phase at the core center.

\section{Kinematics}

\begin{figure}
  \centering      
       \includegraphics[width=3.5in]{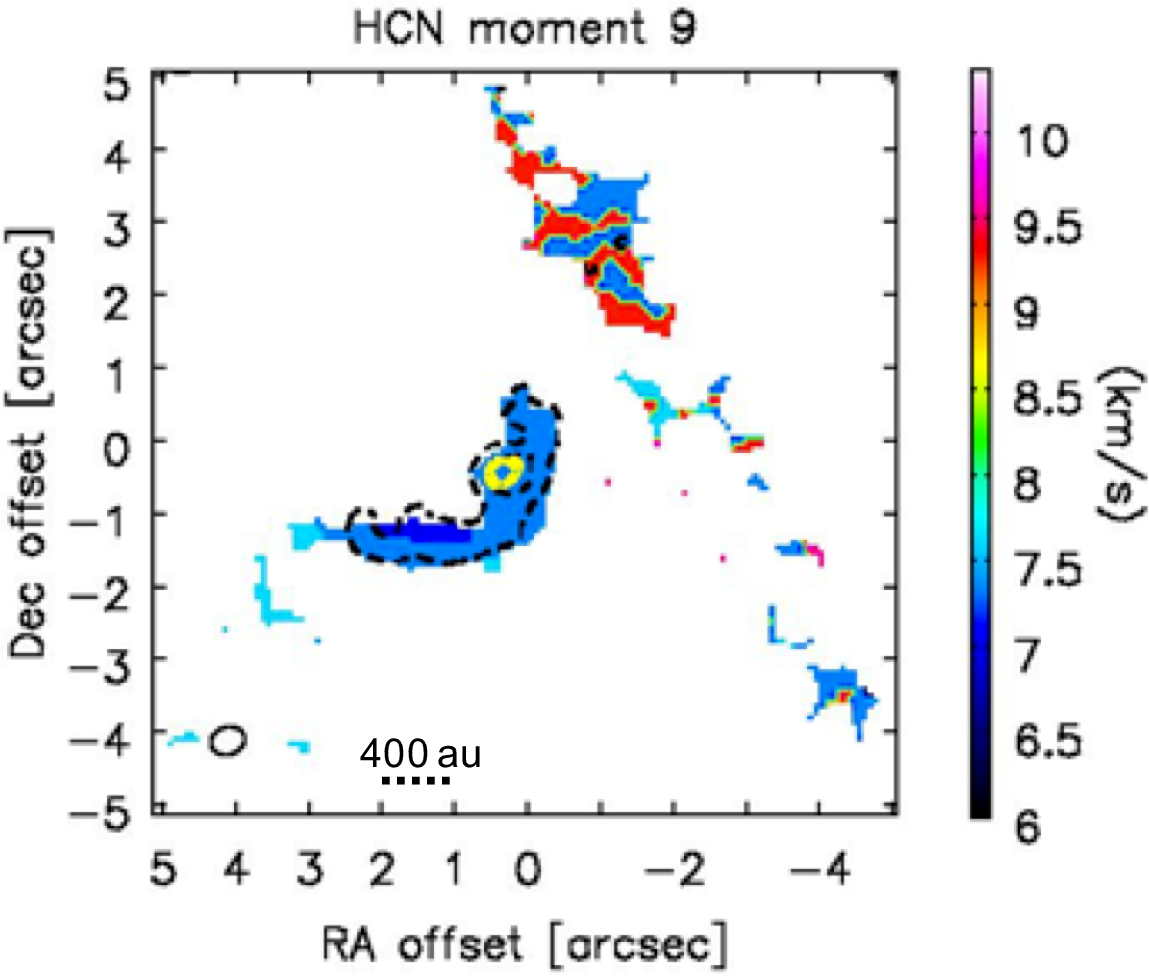}	    
           \caption{The HCN moment 9 map shown in colour raster map. Black dashed contours are the HCN moment 0 map and yellow contours are the continuum map.   }
           \label{mom9}
  \end{figure}

Figure~\ref{mom9} shows the moment 9 (velocity of the maximum value of the spectrum) map in the HCN (3-2) line (all components). With respect to the V$_{lsr}\sim$8.0 km s$^{-1}$ for Ser-emb 16 (Riaz et al. 2024), the emission at the Ser-emb 16 position and the spiral is blue-shifted and does not show any clear velocity gradient with the distance (or radius) over the narrow velocity range of $\sim$7-7.5 km s$^{-1}$. The moment 1 map shows the intensity weighted coordinate, which is traditionally used to get the ``velocity fields'', whereas the moment 9 map shows the velocity of the maximum value of the spectrum. Since there is no velocity gradient seen within the spiral or bar structures, the moment 1 map does not show any velocity fields. The blue- vs. red-shifted velocities can be seen in much more clarity in the moment 9 map than moment 1.

The velocity resolution of our ALMA observations is $\sim$0.3-0.4 km s$^{-1}$. Thus, velocity dispersions of $<$0.6-0.8 km s$^{-1}$ cannot be resolved. This makes it difficult to constrain the position angle and build a position-velocity diagram (PVD) that can help probe the infall and/or Keplerian kinematics for the proto-brown dwarf and the spiral. For a central object mass of 18 M$_{Jup}$ measured for Ser-emb 16, we may find evidence for Keplerian rotation if the velocity spread was $\sim$0.3 km s$^{-1}$ or larger. The non-detection of Keplerian rotation in the bar-like structure is thus mainly due to the limited velocity resolution of the observations. Note that we do not see any jet or outflow for Ser-emb 16, which would support the idea that there is no Keplerian disk in this Class 0 proto-BD system. On the other hand, this also suggests that either the disk has not formed yet or it is still deeply embedded in the pseudo-disk or its size is much smaller than the $\sim$176 au resolution of the observations. Future interferometric observations at a higher velocity and angular resolution can provide more insight into the kinematics of a disk structure in this system.

The moment 9 map also shows a trail of surrounding cloud material towards the north-west of the spiral and extending to the south of it. This trail is associated with a larger scale streamer seen much more brightly in the HCO$^{+}$ (3-2) line emission (Riaz et al. 2024).

\section{Conclusions}

We have explored if the HCN (3-2) line emission could be a useful tracer of the inner Keplerian disk in a proto-brown dwarf system. We find two prominent features in the ALMA HCN (3-2) line observations at $\sim$0.4$\arcsec$ angular resolution towards the proto-brown dwarf, Ser-emb 16. One is a flattened, compact, bar-like structure of size $\sim$50 au, and the other is a spatially extended spiral-shaped structure of size $\sim$1610 au. Radiative transfer modelling indicates that the bar-like structure is tracing the inner disk, albeit confirming the Keplerian kinematics will require at least $\sim$10 times higher velocity resolution than the present observations. The spiral-shaped structure is tracing the pseudo-disk that has been twisted due to core rotation. Combining both results, HCN is found to be a good tracer of the pseudo-disk and the inner disk embedded in it in a proto-brown dwarf system. 

We also present radiative transfer modelling of ALMA HCO$^{+}$ (3-2) line emission observed close to the Ser-emb 16 position. The best model fit indicates that the HCO$^{+}$ is tracing the outer pseudo-disk/envelope regions at $>$200 au. Based on the modelling results, HCN can trace the innermost, densest regions in the proto-brown dwarf where HCO$^{+}$ is depleted from the gas phase.

\section*{Acknowledgements}

B.R. acknowledges funding from the Deutsche Forschungsgemeinschaft (DFG) - Projekt number RI-2919/2-3.

\section{Data Availability}

The data underlying this article are available in the ALMA archives through the ALMA online database.

\title{Supplementary Material}

\appendix
\label{appendix}

\section{Sizes, gas masses and densities}
\label{sizes}

We have used the CASA task {\it imfit} to measure the source size, peak position, position angle, and the line fluxes by fitting one or more elliptical Gaussian components on the continuum and HCN (3-2) line images. The size of Ser-emb 16 measured in the continuum image was used to measure the HCN line fluxes. The integrated and peak HCN (3-2) flux for Ser-emb 16 is 37.88$\pm$5.3 mJy and 31.6$\pm$2.0 mJy beam$^{-1}$. The H$_{2}$ gas mass derived from the HCN column density is 2.8$\pm$0.3 M$_{Jup}$, and the H$_{2}$ gas number density is (1.2$\pm$0.2)$\times$10$^{7}$ cm$^{-3}$.

The size of the spiral, both convolved and deconvolved with the beam, is 3.0$\arcsec \pm$0.6$\arcsec$ along the major axis and 2.1$\arcsec \pm$0.1$\arcsec$ along the minor axis, with a PA of 128.9$\degr$$\pm$4.6$\degr$. The projected size of the spiral is 1610$\pm$264 au. The integrated and peak HCN (3-2) flux for the spiral is 23.4$\pm$3.2 mJy and 23.1$\pm$2.7 mJy beam$^{-1}$. The H$_{2}$ gas mass derived from the HCN column density is 12.2$\pm$0.5 M$_{Jup}$, and the H$_{2}$ gas number density is (0.2$\pm$0.01)$\times$10$^{7}$ cm$^{-3}$.

\section{Comparison of HCN and HCO$^{+}$ maps}
\label{hcop-hcn}

 \begin{figure*}
  \centering      
       \includegraphics[width=1.6in,angle=90]{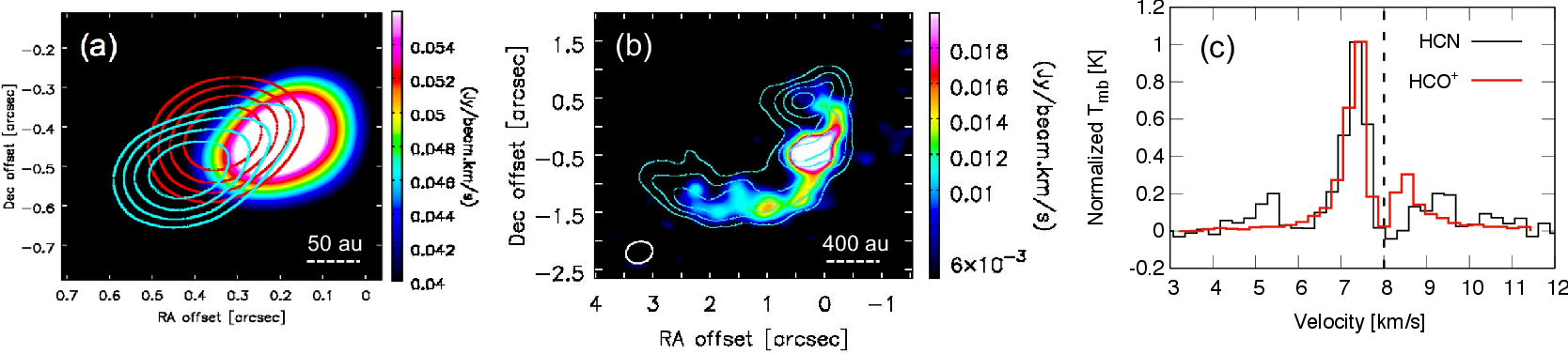}            
           \caption{(a) The observed HCN (colour map) and HCO$^{+}$ (cyan countours) moment 0 maps, with the  continuum emission plotted in red contours. (b) The large-scale spiral seen in the HCN-A (colour map) and HCO$^{+}$ (cyan contours) line emission. (c) An overplot of the HCN (black) and HCO$^{+}$ (red) spectra. The peak intensity has been normalized to one; the normalization factor is 0.82 for HCN and 2.14 for HCO$^{+}$. }
           \label{hcop}
  \end{figure*}

The spectral setup in our ALMA observations included the HCO$^{+}$ (3-2) line. A detailed discussion on the HCO$^{+}$ results are presented in Riaz et al. (2024). Here, we present a comparison of the HCN and the HCO$^{+}$ line maps. Figure~\ref{hcop} shows an overlap of the HCN and HCO$^{+}$ moment 0 maps close to the central Ser-emb 16 source [panel (a)] and a zoomed-out view of the large-scale ($\sim$2000 au) spiral structure seen in both lines [panel (b)]. Close to the proto-brown dwarf position as traced by the continuum emission (red contours), the HCN emission peaks at an offset of $\sim$0.15$\arcsec \sim$66 au while HCO$^{+}$ peaks at an offset of $\sim$0.4$\arcsec \sim$176 au. The HCO$^{+}$ peak is thus twice the distance away from the proto-brown dwarf position than the HCN peak. The HCO$^{+}$ is also more spatially extended, with a spatial extent of $\sim$167 au vs $\sim$123 au for HCN emission close to the central proto-brown dwarf source [Fig.~\ref{hcop}; panel (a)].

If we zoom out over a larger spatial scale [Fig.~\ref{hcop}; panel (b)], then the spiral structure is seen in both lines but the spatial scale (both length and width) is larger by a factor of $\sim$1.5 in HCO$^{+}$ than in HCN. HCN is therefore tracing the more compact and high-density material close to the central proto-brown dwarf source. As discussed in Sect.~4 (main paper) and Sect.~\ref{hcop-model}, line radiative transfer modelling indicates that the HCN peak emission originates from the inner disk, while HCO$^{+}$ emission originates from the outer pseudo-disk/envelope in the proto-brown dwarf. These lines are therefore tracing different physical components, the evidence of which is clearly seen in the position offset and spatial extent in the line maps (Fig.~\ref{hcop}ab).

Interestingly, there is no streamer seen in the HCN (3-2) line map, unlike the impressive large-scale streamer seen in the HCO$^{+}$ (3-2) line (Riaz et al. 2024). The HCO$^{+}$ (3-2) line has an order of magnitude lower critical density of the order of 10$^{6}$ cm$^{-3}$ than the HCN (3-2) line. The surrounding cloud material around an extremely young proto-brown dwarf is expected to be of low density, and thus would be detectable in a comparatively low density tracer such as HCO$^{+}$ than HCN.

Figure~\ref{hcop}c compares the observed HCN (3-2) and HCO$^{+}$ (3-2) spectra collapsed over all velocity channels. The peak intensity has been normalized to one; the normalization factor is 0.82 for HCN and 2.14 for HCO$^{+}$. The HCO$^{+}$ peak intensity is $\sim$2.6 times higher than HCN. The blue-shifted brighter peak in HCO$^{+}$ shows a similar spectral profile as the HCN-A component. Both of these peaks are dominated by the emission from the bright spiral structure (Riaz et al. 2024). The much fainter red-shifted peak in HCO$^{+}$ does not overlap with the HCN satellite components. As discussed in Riaz et al. (2024), this fainter HCO$^{+}$ peak arises from a large-scale streamer that is not observed in the HCN line map.

Note that since the proto-brown dwarf system is viewed at a $\sim$60$\degr$ angle to the line-of-sight, we are seeing one-sided or asymmetric emission from the disk/pseudo-disk/envelope, which could explain why HCN peaks at a slight offset inward of the continuum position while HCO$^{+}$ peaks slightly outward of it (Fig.~\ref{hcop}a).


\section{HCO$^{+}$ Modelling}
\label{hcop-model}

\begin{figure*}
  \centering      
       \includegraphics[width=1.8in,angle=90]{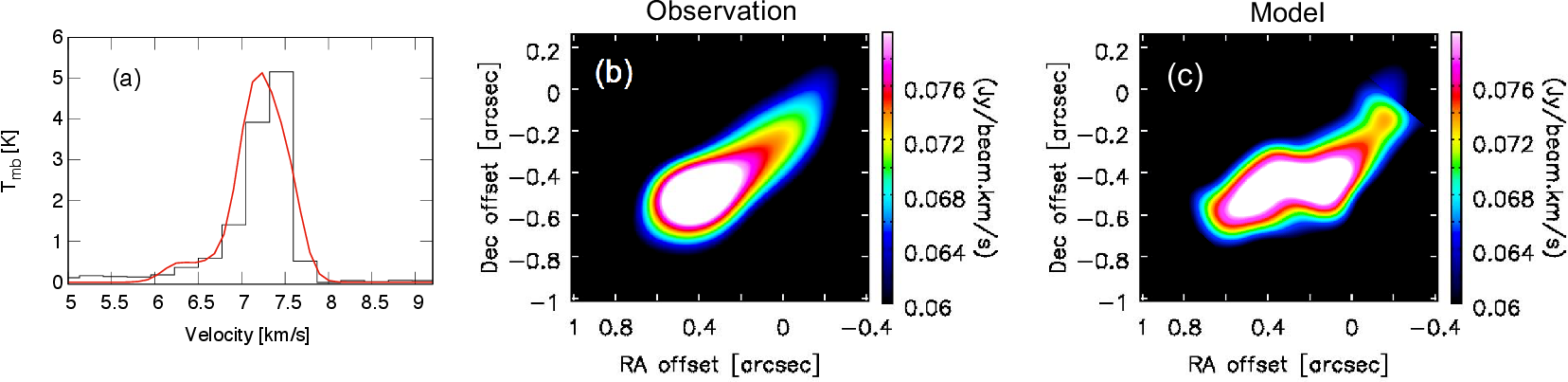}     \\	\vspace{0.2in}
       \includegraphics[width=2in,angle=0]{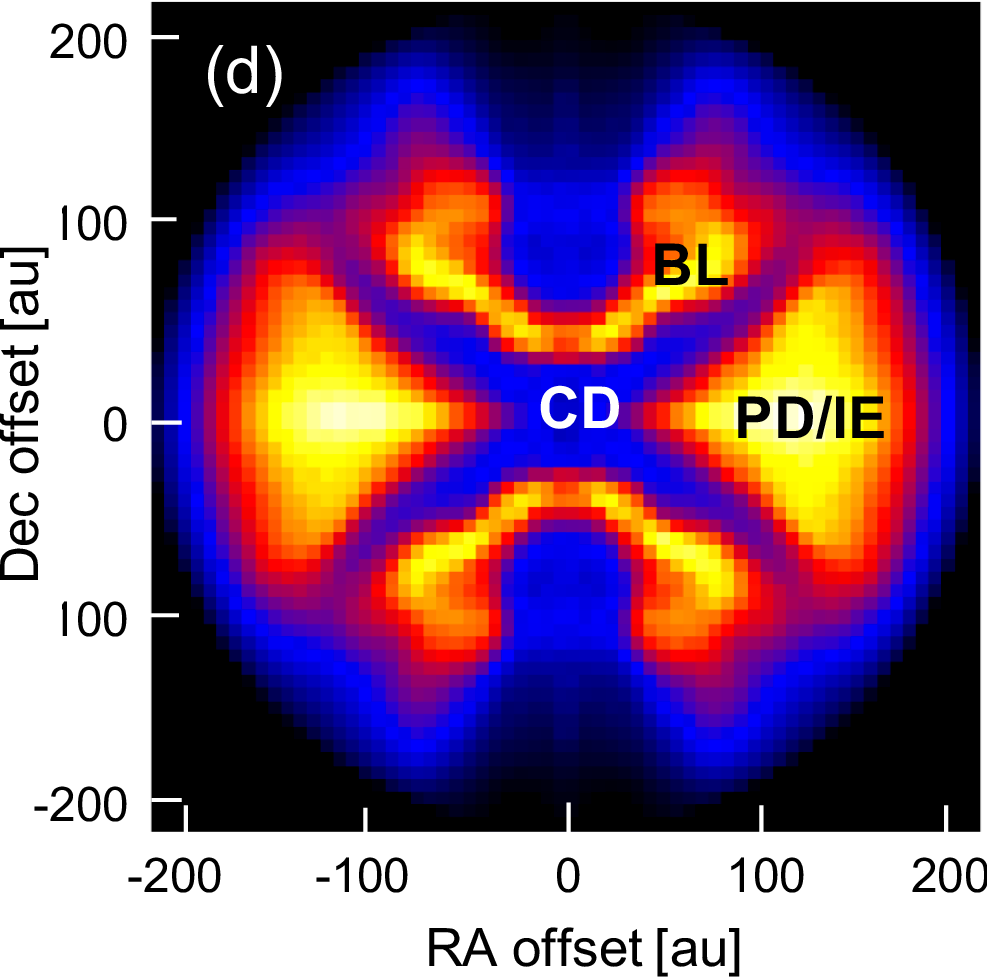}                   
           \caption{Results from radiative transfer modelling of the HCO$^{+}$ emission close to the proto-brown dwarf. (a) Observed (black) and best model-fit (red) spectrum. (b) Observed moment 0 map. (c) Model moment 0 map produced from the best line fit. (d) Model moment 0 map at an edge-on inclination. The labels `CD', `PD/IE', and `BL' indicate central depletion, pseudo-disk/inner envelope, and boundary layer, respectively. }
           \label{hcop-fit}
  \end{figure*}

A detailed modelling of the large-scale spiral structure seen in the HCO$^{+}$ (3-2) line is presented in Riaz et al. (2024). Here we present results from modelling the HCO$^{+}$ emission close to the central proto-brown dwarf (Fig.~\ref{hcop}a). We used the same modified physical model by Machida et al. (2020) as used for the HCN data. The best model fit is shown in Fig.~\ref{hcop-fit}a. The model moment 0 map produced from the best line model fit (Fig.~\ref{hcop-fit}c) shows a twist around $\sim$0.3$\arcsec$ or $\sim$130 au, which is likely a line-of-sight effect.

Fig.~\ref{hcop-fit}d shows the model image at an edge-on inclination. The peak in the HCO$^{+}$ emission is seen in the outer pseudo-disk/envelope regions at $>$200 au (labelled `PD/IE'), and in the boundary layers (labelled `BL') between the high-velocity jet and the inner pseudo-disk/envelope regions. The model also predicts a significant decline in the HCO$^{+}$ intensity towards the central regions (labelled `CD'). This suggests that the HCO$^{+}$ molecule is depleted in the inner, dense regions of the proto-brown dwarf.

\bsp	
\label{lastpage}
\end{document}